\pdfoutput=1  

\documentclass[11pt]{article}
\usepackage{rotating}
\makeatletter
\@addtoreset{equation}{section}
\makeatother

\def\dalemb#1#2{{\vbox{\hrule height .#2pt
        \hbox{\vrule width.#2pt height#1pt \kern#1pt
                \vrule width.#2pt}
        \hrule height.#2pt}}}

  \let\q=\theta  
  \let\n=\nu

\let\C=\Chi      
\let\la=\label  
 \def\bd{\begin{document}} \def\ed{\end{document}}
\def\ds{\documentstyle} \let\fr=\frac \let\bl=\bigl \let\br=\bigr
\let\Br=\Bigr \let\Bl=\Bigl 
\let\bm=\bibitem
\let\na=\nabla
\let\pa=\partial \let\ov=\overline
\def\ie{{\it i.e.\ }} 
\newcommand{\be}{\begin{equation}} 
\newcommand{\ee}{\end{equation}} 
\def\ba{\begin{array}}
\def\ea{\end{array}}
\def\ft#1#2{{\textstyle{{\scriptstyle #1}\over {\scriptstyle #2}}}}
\def\fft#1#2{{#1 \over #2}}
\def\del{\partial}
\def\sst#1{{\scriptscriptstyle #1}}
\def\oneone{\rlap 1\mkern4mu{\rm l}}
\def\td{\tilde}
\def\wtd{\widetilde}
\def\im{{\rm i}}
\def\bog{Bogomol'nyi\ }
\def\q{{\tilde q}}
\def\hast{{\hat\ast}}
\def\0{{\sst{(0)}}}
\def\1{{\sst{(1)}}}
\def\2{{\sst{(2)}}}
\def\3{{\sst{(3)}}}
\def\4{{\sst{(4)}}}
\def\5{{\sst{(5)}}}
\def\6{{\sst{(6)}}}
\def\7{{\sst{(7)}}}
\def\8{{\sst{(8)}}}
\def\n{{\sst{(n)}}}
\newcommand{\w}[1]{\\[0.#1cm]}
\def\hA{\hat{\cal A}}
\def\ns{{\sst {\rm NS}}}
\def\rr{{\sst {\rm RR}}}
\def\tH{{\widetilde H}}
\def\tB{{\widetilde B}}
\def\cA{{\cal A}}
\def\cF{{\cal F}}
\def\tF{{\wtd F}}
\def\v{{{\cal V}}}
\def\Z{\rlap{\sf Z}\mkern3mu{\sf Z}}
\def\ep{{\epsilon}}
\def\IIA{{\rm IIA}}
\def\IIB{{\rm IIB}}
\def\ads{{\rm AdS}}
\def\R{\rlap{\rm I}\mkern3mu{\rm R}}
\def\ua{\underline{\alpha}}
\def\ub{\underline{\phantom{\alpha}}\!\!\!\beta}
\def\uc{\underline{\phantom{\alpha}}\!\!\!\gamma}
\def\um{\underline{\mu}}
\def\ud{\underline\delta}
\def\ue{\underline\epsilon}
\def\una{\underline a}
\def\unA{\underline A}
\def\unb{\underline b}
\def\unB{\underline B}
\def\unc{\underline c}
\def\unC{\underline C}
\def\und{\underline d}
\def\unD{\underline D}
\def\une{\underline e}
\def\unE{\underline E}
\def\unf{\underline{\phantom{e}}\!\!\!\! f}
\def\unF{\underline F}
\def\ung{\underline g}
\def\unm{\underline m}
\def\unM{\underline M}
\def\unn{\underline n}
\def\unN{\underline N}
\def\unp{\underline{\phantom{a}}\!\!\! p}
\def\unP{\underline P}
\def\unH{\underline{H}}
\def\unF{\underline{F}}
\def\unT{\underline{T}}
\def\ovA{\overline{A}}
\def\ovB{\overline{B}}
\def\uC{{\underline C}}
\def\ns{\normalsize}
\def\vs{\vspace{-0.25cm}}
\def\se{\;\;=\;\;}
\def\de{\;\;:=\;\;}
\def\cF{{\cal F}}
\def\cH{{\cal H}}
\def\cK{{\cal K}}

\def\atwo{\alpha_{2}}
\def\aone{\alpha_{1}}
\def\afive{\alpha_{5}}
\def\ap{\alpha_p}
\def\azero{\alpha_o}
\def\afour{\alpha_{4}}
\def\appt{\alpha_{p+2}}
\def\apmo{\alpha_{p-1}}

\def\cE{{\cal E}}
\def\tr{{\rm tr}}
\def\bC{{\bar \C}}

\newcommand{\bea}{\begin{eqnarray}} 
\newcommand{\eea}{\end{eqnarray}} 
\newcommand{\ra}{\rightarrow}
\newcommand{\Tr}{{\rm Tr} } 
\newcommand{\tamphys}{\it Department of Physics, Imperial College London, Prince Consort Road,
London SW7 2BZ}

\newcommand{\auth}{M. J. Duff\footnote{mduff@imperial.ac.uk}}

\begin{document}
\begin{flushright}
\end{flushright}

\vspace{24pt}

\begin{center}
{ \large {\bf Near-horizon brane-scan revived}}

\vspace{36pt}

\auth

\vspace{10pt}

{\tamphys}

\vspace{44pt}

\underline{ABSTRACT}

\end{center}

In 1987 two versions of the brane-scan of $D$-dimensional super $p$-branes were put forward. The first pinpointed those $(p,D)$ slots consistent with kappa-symmetric Green-Schwarz type actions; the second generalized the {\it membrane at the end of the universe} idea to all those superconformal groups describing $p$-branes on the boundary of $AdS_{p+2} \times S^{D-p-2}$.  Although the second version predicted $D3$ and $M5$ branes in addition to those of the first, it came unstuck because the 1/2 BPS solitonic branes failed to exhibit the required symmetry enhancement in the near-horizon limit, except in the non-dilatonic cases $(p=2,D=11)$, $(p=3,D=10)$ and $(p=5,D=11)$. Just recently, however, it has been argued that the fundamental $D=10$ heterotic string does indeed display a near-horizon enhancement to $OSp(8|2)$ as predicted by the brane-scan, provided $\alpha'$ corrections are taken into account. If this logic could be extended to the other strings and branes, it would resolve this 21-year-old paradox and provide a wealth of new AdS/CFT dualities, which we tabulate.
{\vfill\leftline{}\vfill

\pagebreak
\setcounter{page}{1}

\tableofcontents
\addtocontents{toc}{\protect\setcounter{tocdepth}{2}}
\newpage
\section{\bf Two brane-scans}
\la{intro}

\indent

In 1987 two versions of the brane-scan of $D$-dimensional super $p$-branes were put forward. The first by Achucarro, Evans, Townsend and Wiltshire \cite{AETW} pinpointed those twelve $(p,D)$ slots consistent with kappa-symmetric Green-Schwarz \cite{Green:1983wt} type actions for $p\geq 1$ . The result is shown in Table \ref{kappascan}.   In the early eighties Green and Schwarz had shown that spacetime supersymmetry allows classical superstrings moving in spacetime dimensions
$3,4,6$ and $10$, with $D=10$ case being anomaly-free.  It was now realized, however, that these $1$-branes in $D=3,4,6$ and $10$ should now be viewed as the endpoints of four sequences of $p$-branes. Moving diagonally down the brane-scan corresponds to a simultaneous dimensional reduction of spacetime and worldvolume \cite{Duff:1987bx}.  Of course some of these $D$ dimensions could be compactified, in which case the double dimensional reduction may be interpreted as wrapping the brane around the compactified directions.  We shall return to compactifications in Sections \ref{Generalizations} and \ref{new}. Note also that these are all 1/2 BPS branes. Intersecting branes with less supersymmetry are discussed in Section \ref{Generalizations}.

\begin{table}
\begin{center}
\begin{tabular}{ccccccccccccccc}
~&$D\uparrow$&&&&&&&&&&&~\\
~&11&.&&&S&&&&&&&&~\\
~&10&.&&S&&&&S&&&&&~\\
~&9&.&&&&&S&&&&&&~\\
~&8&.&&&&S&&&&&&&~\\
~&7&.&&&S&&&&&&&&~\\
~&6&.&&S&&S&&&&&&&~\\
~&5&.&&&S&&&&&&&&~\\
~&4&.&&S&S&&&&&&&&~\\
~&3&.&&S&&&&&&&&&~\\
~&2&.&&&&&&&&&&&~\\
~&1&.&~&~&~&~&~&~&~&~&~&~&~\\
~&0&.&.&.&.&.&.&.&.&.&.&.&.~\\
~&~&0&1&2&3&4&5&6&7&8&9&10&11&$d\rightarrow$
\end{tabular}
\end{center}
\caption{ $p$-branes described by Green-Schwarz actions, with scalar worldvolume supermultiplets}
\la{kappascan}
\end{table}

The second brane-scan by Blencowe and the author \cite{BD} generalized the {\it membrane at the end of the universe} idea \cite{Fifteen, BDPS} to arbitrary $p$-branes with $p\geq 1$ by selecting those superconformal groups in Nahm's classification \cite{Nahm} (listed in appendix \ref{ads}) with bosonic subgroups $SO(p+1,2) \times SO(D-p-1)$ describing $p$-branes on the boundary of  $AdS_{p+2} \times S^{D-p-2}$, as shown in Table \ref{horscan}. 
\begin{table}
\begin{center}
\begin{tabular}{lcl}
\hline
Supergroup&AdS Dimension&$p$\\
&&\\
\hline
$OSp(n_+|2) \times OSp(n_-|2)$&3& 1 \\
$OSp(N|4)$& 4& 2 \\
$SU(2,2|N)$&5&3\\
$F^2(4)$&6&4 \\
$OSp(8^*|N)$&7&5\\
\hline
\end{tabular}
\caption{ Supergroups admitting $p$-branes on the boundary of $AdS_{p+2} \times S^{D-p-2}$.}
\label{horscan}
\end{center}
\end{table}
In each case the boundary CFT is described by the corresponding singleton (scalar), doubleton (scalar or vector) or tripleton (scalar or tensor) supermultiplet\footnote{Our nomenclature is based on the rank of $AdS_{p+2}$ and differs from that of G\"{u}naydin \cite{Gunaydin1}.}.  The supersingleton lagrangian and transformation rules were also spelled out explicitly in this paper. The resulting brane-scans are shown in  Tables \ref{scalar}, \ref{vector} and \ref{tensor}, where $d=p+1$.  The number of dimensions transverse to the brane, $D-d$, equals the number of scalars in the singleton, doubleton or tripleton  supermultiplet, as shown in Table \ref{fields}.  Once again at this stage we are considering only 1/2 BPS branes in uncompactified spacetimes.  The two factors appearing in the $d=2$ case is simply a reflection of the ability of strings to have right and left movers.  For brevity, we have written the Type II assignments in Table \ref{scalar}, but more generally we could have $OSp(n_+|2) \times OSp(n_-|2)$ where $n_+$ and $n_-$ are the number of left and right supersymmetries \cite{Gunaydin:1986cs}.
\begin{table}
\begin{center}
\begin{tabular}{cccc}
$D$&{\bf Supergroup}&{\bf Supermultiplet}&{\bf Field~content}\\
10&$OSp(8|2)$&$((n_+,n_-)=(8,0),d=2)$~singleton&8~spinors, 8~scalars\\
6&$OSp(4|2)$&$((n_+,n_-)=(4,0),d=2)$~singleton&4~spinors, 4~scalars,\\
4&$OSp(2|2)$&$((n_+,n_-)=(2,0),d=2)$~singleton&2~spinors, 2~scalars\\
3&$OSp(1|2)$&$((n_+,n_-)=(1,0),d=2)$~singleton&1~spinor, 1~scalar\\
11&$OSp(8|4)$&$(n=8,d=3)$~singleton&8~spinors, 8~scalars\\
7&$OSp(4|4)$&$(n=4,d=3)$~singleton&4~spinors, 4~scalars\\
5&$OSp(2|4)$&$(n=2,d=3)$~singleton&2~spinors, 2~scalars\\
4&$OSp(1|4)$&$(n=1,d=3)$~singleton&1~spinor, 1~scalar\\
8&$SU(2,2|2)$&$(n=2,d=4)$~doubleton&2~spinors, 4~scalars\\
6&$SU(2,2|1)$&$(n=1,d=4)$~doubleton&1~spinor, 2~scalars\\
9&$F^2(4)$&$(n=2,d=5)$~doubleton&2~spinors,4~scalars\\
10&$OSp(8^*|2)$&$((n_{+}¥,n_{-}¥)=(1,0),d=6)$~tripleton&2~spinors, 4~scalars\\
&&&\\
10&$SU(2,2|4)$&$(n=4,d=4)$~doubleton&1~vector, 4~spinors, 6~scalars\\
6&$SU(2,2|2)$&$(n=2,d=4)$~doubleton&1~vector, 2~spinors, 2~scalars\\
4&$SU(2,2|1)$&$(n=1,d=4)$~doubleton&1~vector, 1~spinor\\
&&&\\
11&$OSp(8^*|4)$&$((n_{+}¥,n_{-}¥)=((2,0),d=6)$~tripleton&1~chiral~2-form, 4~spinors, 5~scalars\\
7&$OSp(8^*|2)$&$((n_{+}¥,n_{-}¥)=((1,0),d=6)$~tripleton&1~chiral~2-form, 2~spinors, 1~scalar\\
\end{tabular}
\end{center}
\medskip
\caption{Superconformal groups and their singleton, doubleton 
and tripleton representations. $D=d$ + number of scalars.}
\label{fields}
\end{table}

Note that Table \ref{scalar} reproduces the same twelve points as Table \ref{kappascan}. However, Tables \ref{vector} and \ref{tensor} predicted new branes including what would later be called the $D3$ and $M5$ branes, which at the time were more mysterious.

\begin{table}
\begin{center}
\begin{tabular}{cccccccccc}
~&D$\uparrow$&&&&&&\\
~&11&.&~~~~~~~~~~~~~~~~~~&&${\bf OSp(8|4)}$&&&\\
~&10&.&~~~~~~~~&$[OSp(8|2)]^2$&&&&$OSp(8^*|2)$\\
~&~9&.&~~~~~~~~&&&&$F^2(4)$&\\
~&~8&.&~~~~~~~~&&&$SU(2,2|2)$&&\\
~&~7&.&~~~~~~~~&&$OSp(4|4)$&&&\\
~&~6&.&~~~~~~~~&$[OSp(4|2)]^2$&~&$SU(2,2|1)$&&\\
~&~5&.&~~~~~~~~&&$OSp(2|4)$&&&\\
~&~4&.&~~~~~~~~&$[OSp(2|2)]^2$&$OSp(1|4)$&&&\\
~&~3&.&~~~~~~~~&$[OSp(1|2)]^2$&~&&&\\
~&~2&.&~~~~~~~~&&&&&&\\
~&~1&.&~~~~~~~~&~&~&~&~&\\
~&~0&.&.&.&.&.&.&.\\
~&~~&0&1&2&3&4&5&6&d$\rightarrow$
\end{tabular}
\end{center}
\bigskip
\caption{The brane-scan of superconformal groups admitting 
scalar supermultiplets: singletons ($p=1,2$), doubletons ($p=3,4)$ and tripletons ($p=5$).}
\la{scalar}
\end{table}
\begin{table}
\begin{center}
\begin{tabular}{cccccccccc}
~&D$\uparrow$&&&&&&\\
~&11&.&~~~~~~&~~~~~~~~~~~~&&~~~~~~~~~~~~&~~~~~~~~~~~~&\\
~&10&.&~~~~~~&&&${\bf SU(2,2|4)}$&&\\
~&9&.&~~~~~~&&&&&\\
~&8&.&~~~~~~&&&&&\\
~&7&.&~~~~~~&&&&&\\
~&6&.&~~~~~~&&&$SU(2,2|2)$&&\\
~&5&.&~~~~~~&&&&&\\
~&4&.&~~~~~~&&&$SU(2,2|1)$&&\\
~&3&.&~~~~~~&&&&&\\
~&2&.&~~~~~~&&&&&&\\
~&1&.&~~~~~~&~&~&~&~&\\
~&0&.&~~~~~~.&.&.&.&.&.\\
~&~&0&~~~~~~1&2&3&4&5&6&d$\rightarrow$
\end{tabular}
\end{center}
\bigskip
\caption{The brane-scan of superconformal groups 
admitting vector supermultiplets: doubletons ($p=3$).}
\label{vector}
\end{table}
\begin{table}
\begin{center}
\begin{tabular}{cccccccccc}
~&D$\uparrow$&&&&&&\\
~&11&.&~~~~~~&~~~~~~~~~~~~&~~~~~~~~~~~~&~~~~~~~~~~~~&~~~~~~~~~~~~&
${\bf OSp(8^*|4)}$\\
~&10&.&~~~~~~&&&&~&\\
~&9&.&~~~~~~&&&&&\\
~&8&.&~~~~~~&&&&&\\
~&7&.&~~~~~~&&&&&$OSp(8^*|2)$\\
~&6&.&~~~~~~&&&&~&~\\
~&5&.&~~~~~~&&&&&\\
~&4&.&~~~~~~&&&~&&\\
~&3&.&~~~~~~&&~&&&\\
~&2&.&~~~~~~&&&&&&\\
~&1&.&~~~~~~&~&~&~&~&\\
~&0&.&~~~~~~.&.&.&.&.&.\\
~&~&0&~~~~~~1&2&3&4&5&6&d$\rightarrow$
\end{tabular}
\end{center}
\bigskip
\caption{The brane-scan of superconformal groups 
admitting tensor supermultiplets: tripletons ($p=5$).}
\label{tensor}
\end{table}

In early 1988, Nicolai, Sezgin and Tanii \cite{NST} independently put forward the same generalization of the {\it membrane at the end of the universe} idea, spelling out the doubleton and tripleton lagrangian and transformation rules, in addition to the singleton.
However, by insisting on only scalar supermultiplets their list corresponded to the branes of Table \ref{scalar}, but not those of Tables \ref{vector} and \ref{tensor}.  In this case, as they note, the spheres happen to be the parallelizable ones $S^1$, $S^3$ and $S^7$.

Notwithstanding the agreement between the $(p,D)$ slots in Table \ref{scalar} and Table \ref{kappascan} and notwithstanding the prediction of new branes in Tables \ref{vector} and \ref{tensor}, including $D3$ and $M5$, the {\it end of the universe} brane-scans suffered from the following problem:
Although the corresponding $D$-dimensional supergravity theories all admit compactifications to  $AdS_{p+2} \times S^{D-p-2}$, the resulting bosonic symmetry is not in general  $SO(p+1,2) \times SO(D-p-1)$ and hence the resulting supergroups are not those of Table \ref {horscan}. For example, Type I supergravity in $D=10$ admits solutions of the form $AdS_7 \times S^3$ and $AdS_3 \times S^7$ but there is a non-constant dilaton whose gradient acts as a conformal AdS Killing vector \cite{DTV}. So the symmetries are not $OSp(8|2)$ or $OSp(8^*|2)$ as required by the brane-scan. Exceptions to this rule are provided by the non-dilatonic solutions $(p=2,D=11)$, $(p=3,D=10)$ and $(p=5,D=11)$ denoted in boldface in Tables \ref{scalar}, \ref{vector} and \ref{tensor}.

This problem raised it head once more with the arrival of 1/2 BPS solitonic $p$-branes \cite{String}  which, with the above exceptions \cite{GT} , failed to exhibit the required symmetry enhancement in the near-horizon limit \cite{DGT},  as described in Section \ref{Aristocrat}.

Just recently, however, it has been argued by  Dabholkar and Murthy \cite{Dabholkar:2007gp} and by Lapan, Simons and Strominger \cite{Lapan:2007jx}, that the fundamental $D=10$ heterotic string \cite{Dabholkar:1990yf,Duff:1991sz} does indeed display a near-horizon enhancement to $OSp(8|2)$ as predicted by the brane-scan, provided $\alpha'$ corrections are taken into account. See also Johnson \cite{Johnson} and Kraus, Larsen and Shah \cite{Kraus:2007vu}. This is taken up in Section \ref{new}. 

\section{The near-horizon geometry problem}
\la{Aristocrat}

$AdS$ emerges in the near-horizon geometry of black
$p$-brane solutions \cite{GT,DGT,String} in $D$ dimensions. The dual
brane, with worldvolume dimension ${\tilde d}=D -d-2$, interpolates
between $D$-dimensional Minkowski space $M_{D}¥$ and $AdS_{{\tilde d}+1}
\times
S^{d+1}$ (or $M_{{\tilde d}+1}\times S^{3}$ if $d=2$).  To see this, 
we recall that such branes arise generically as  
solitons of the following action \cite{lublack}:
\be
I=\frac{1}{2\kappa_{D}¥^{2}}¥ \int d^{D}¥x \sqrt{-g}\left[R-\frac{1}{2} 
(\partial 
\phi)^{2}¥-\frac{1}{2(d+1)!}e^{-\alpha \phi} F_{d+1}{}^{2}\right]
\ee
where $F_{d+1}¥$ is the field strength of a $d$-form potential 
$A_{d}¥ $ and $\alpha$ is the constant
\be
\alpha^{2}¥=4-\frac{2d \tilde d}{d+\tilde d}
\la{alpha}
\ee
Written in terms of the 
$(d-1)$-brane sigma-model metric $e^{-{\alpha/d} \phi}¥g_{MN}¥$, the 
solutions are \cite{lublack,String}
\[
ds^{2} =H^{\frac{2-d}{d}}¥dx^{\mu}¥dx_{\mu}¥+
H^{2/d}¥(dy^{2}¥+y^{2}¥d\Omega_{d+1}¥{}^{2})¥
\]
\[
e^{2\phi }¥=H^{\alpha}¥
\]
\be
F_{d+1}¥=dL^{d}¥\epsilon_{d+1}¥
\la{branesolution}
\ee
where 
\be
H=1+\frac{L^{d}}{y^{d}}
\ee
For a stack of $N$ singly charged branes $L^{d}¥=Nb^{d}¥$ and the near 
horizon, or large $N$, geometry corresponds to 
\be
ds^{2}¥\sim({\frac{y}{L}})^{2-d}¥dx^{\mu}¥dx_{\mu}+\frac{L^{2}¥}{y^{2}}¥dy^{2}¥
+L^{2}¥d\Omega_{d+1}^{2}¥¥
\ee
Or, defining the new coordinate
\be
y=Le^{\zeta/L}
\ee
we get
\[
ds^{2}\sim e^{\frac{2-d}{L}\zeta}dx^{\mu}¥dx_{\mu} +d\zeta^{2}¥
+L^{2}¥d\Omega_{d+1}{}^{2}
\]
\[
\phi \sim   \frac{d\alpha}{2L}\zeta
\]
\be
F_{d+1}¥ \sim dL^{d}¥\epsilon_{d+1}¥
\la{nearhorizon}
\ee
Thus for $d\neq 2$ the near-horizon geometry is $AdS_{\tilde d +1}¥ \times 
S^{d+1}¥$. Note, however, that the gradient of the dilaton is 
generically non-zero and plays the role of a conformal Killing vector on 
$AdS_{\tilde d +1}$. Consequently, there is no enhancement of 
symmetry in the near-horizon limit. The unbroken supersymmetry remains 
one-half and the bosonic symmetry remains $P_{\tilde d}¥\times 
SO(d+2)$.
(If $d=2$, then (\ref{nearhorizon}) reduces to
\[
ds^{2} \sim dx^{\mu}¥dx_{\mu} + d\zeta^{2}+L^{2}¥d\Omega_{3}{}^{2}  
\]
\[
\phi \sim \frac{\alpha}{L}\zeta
\]
\be
F_{3}¥ \sim 2L^{2}¥\epsilon_{3}¥
\ee
which is $M_{\tilde d+1}¥\times S^{3}¥$, with a linear dilaton 
vacuum. The bosonic symmetry remains $P_{\tilde d}¥\times 
SO(4)$.)

Of particular interest are the ($\alpha=0$) subset of solitons for which 
the dilaton is zero or constant: the {\it non-dilatonic $p$-branes}. From 
(\ref{alpha}) we see that for branes with one kind of charge there are only 3 
cases:
\[
D=11: d=6, \tilde d=3
\]
\[
D=10: d=4, \tilde d=4
\] 
\[
D=11: d=3, \tilde d=6
\]
which are precisely the three cases of M2 \cite{Duff:1990xz}, D3 \cite{HS,DLgauge} and M5 \cite{Gueven} that occupied privileged positions in Tables \ref{scalar}, \ref{vector} and \ref{tensor}.  Then the near-horizon geometry coincides with the $AdS_{\tilde d +1}  \times S^{d+1}$ non-dilatonic maximally symmetric compactifications of the corresponding supergravities. The supersymmetry doubles and the bosonic symmetry is also enhanced to $SO(\tilde d,2) \times SO(d+2)$. Thus the total symmetry is given by the conformal supergroups $OSp(8|4)$, $SU(2,2|4)$ and $OSp(8^*|4)$, respectively.

Thus we see that not all branes are created equal. A {\it $p$-brane 
aristocracy} obtains whose members are those branes whose near-horizon
geometries have as their symmetry the conformal supergroups. As an example
of a plebeian brane we can consider the ten-dimensional  superstring:
\[
D=10: d=6, \tilde d=2
\]
whose near-horizon geometry is $AdS_{3} \times S^{7}$ but with a 
non-trivial dilaton which does not have 
the conformal group $OSp(8|2)$ as its symmetry, even though 
this group appears in the $(D=10, \tilde d=2)$ slot on the scalar 
brane-scan of Table \ref{scalar}.  Such mismatches were the primary reason that the near-horizon brane-scan slipped into oblivion, but we shall consider its revival in Section \ref{new}.

\section{Generalizations}
\la{Generalizations}

\subsection{New kappa-symmetric brane-scan}
\la{kappa}

An equivalent way to arrive at the Green-Schwarz brane-scan of Table \ref{kappascan} is to list
all scalar supermultiplets in $d \geq 2$ dimensions and to interpret the dimension of the target space, $D$, by
\be
{D - d =~{\rm number~of~scalars}.}
\ee
In particular, we can understand $d_{{\rm max}} = 6$ from this point of view since
this is the upper limit for scalar supermultiplets.  However, with the discovery in 1990 of Type II $p$-brane solitons \cite{CHS1,CHS2,HS,DLgauge,Luscan}, and with the discovery in 1992 of the $M$-theory 5-brane \cite{Gueven}, vector and tensor multiplets were also seen to play a role.   These developments gave rise to the new kappa-symmetric brane-scan \cite{Luscan,String} of Table 
\ref{branescannew}.

In particular, the worldvolume fields of the self-dual Type IIB super 3-brane were shown to
be described by an $(n=4,d=4)$ gauge theory \cite{DLgauge}, which on the
boundary of $AdS_5$ is just the doubleton supermultiplet of the
superconformal group $SU(2,2|4)$! Similarly, the worldvolume fields of the $M$-theory 5-brane were shown to be described by an
$((n_+,n_-)=(2,0),d=6)$ multiplet with a chiral $2$-form, $8$ spinors and 
$5$ scalars \cite{GT}, which on the boundary of $AdS_7$ is just the tripleton 
supermultiplet of the superconformal group $OSp(8^*|4)$! (These zero modes are the 
same as those of the Type IIA 5-brane, found previously in \cite{CHS1,CHS2}). 

So although the confirmation of the Type II $3$-brane and M-theory $5$-brane should have provided a feather in the cap of the near-horizon brane-scan, the failure of all the other branes, except the M-theory $2$-brane, to exhibit a near-horizon symmetry enhancement still blighted its success.

With the inclusion of branes with vector and tensor supermultiplets on 
their worldvolume, another curiosity arises. Whereas the near-horizon brane-scan of Table \ref{scalar} exhausts all the scalar branes and the near-horizon brane-scan 
of Table \ref{tensor} exhausts all the tensor branes, the near-horizon brane 
scan of Table \ref{vector} is only a subset of all the vector branes \cite{super}. The Type IIB $3$-brane is special because gauge theories are conformal only in $d=4$.  Branes with vectors on their worldvolume in $d\neq 4$ are doomed to remain forever plebeian\footnote{Not counting $d=3$ Chern-Simons models.}.

\begin{table}
\begin{center}
\begin{tabular}{ccccccccccccccc}
~&$D\uparrow$&&&&&&&&&&&~\\
~&11&.&~&&S&&&T&&&&&~\\
~&10&.&V&S/V&V&V&V&S/V&V&V&V&V&~\\
~&9&.&S&&&&S&&&&&&~\\
~&8&.&~&&&S&&&&&&&~\\
~&7&.&~&&S&&&T&&&&&~\\
~&6&.&V&S/V&V&S/V&V&V&&&&&~\\
~&5&.&S&&S&&&&&&&&~\\
~&4&.&V&S/V&S/V&V&&&&&&&~\\
~&3&.&S/V&S/V&V&&&&&&&&~\\
~&2&.&S&&&&&&&&&&~\\
~&1&.&~&~&~&~&~&~&~&~&~&~&~\\
~&0&.&.&.&.&.&.&.&.&.&.&.&.~\\
~&~&0&1&2&3&4&5&6&7&8&9&10&11&$d\rightarrow$
\end{tabular}
\end{center}
\caption{ The new kappa-symmetric brane-scan, where $S$, $V$ and $T$ denote scalar, vector and
antisymmetric tensor multiplets.}
\la{branescannew}
\end{table}

\subsection{Intersecting branes}
\la{intersecting}

So far we have considered only single 1/2 BPS branes but intersecting branes with less supersymmetry can also exhibit AdS near-horizon geometry. For bound states of branes with $M$ kinds of charge, the constant $\alpha$ of Section \ref{Aristocrat} gets replaced by \cite{DR2,DR,KKLP}
\be
\alpha^{2}¥=\frac{4}{M}-\frac{2d \tilde d}{d+\tilde d}
\la{alpha2}
\ee
A non-dilatonic solution ($\alpha$=0) occurs for $M=2$: 
\[
D=6: d=2, \tilde d=2 
\]
which is just the dyonic string \cite{Rahmfeld}, of which the self-dual 
string 
\cite{lublack} is a special case, whose near-horizon geometry is 
$AdS_{3}¥\times S^{3}$. For $M=3$ we have
\[
D=5: d=2, \tilde d=1
\]
which is the 3-charge black hole \cite{T}, whose near-horizon geometry is 
$AdS_{2} \times S^{3}$, and
\[
D=5: d=1, \tilde d=2
\]
which is the 3-charge string \cite{T} whose near-horizon geometry is 
$AdS_{3} \times S^{2}$. For $M=4$ we have
\[
D=4: d=1, \tilde d=1
\]
which is the 4-charge black hole \cite{CT1,CT2}, of which the 
Reissner-Nordstr\"{o}m  
solution is a special case \cite{DR2}, and whose near-horizon geometry 
is $AdS_{2} \times S^{2}¥$ \cite{FG}.

\subsection{Other signatures}
\la{signatures}

The {\it $p$-brane at the end of the universe} idea  may also be applied in different spacetime signatures. See \cite{BD2,Hull,Batrachenko:2002pu} for a discussion of the corresponding superconformal groups.

\subsection{Compactifications}
\la{Compactifications}

So far we have considered only uncompactified branes with near horizon geometry $AdS_{p+2} \times S^{D-p-2} $.  One might also consider critical (or non-critical) strings and branes compactified on $T^k$, for example. Then the brane-scan would suggest near-horizon geometries of the form $AdS_{p+2} \times S^{D-k- p-2} \times T^k$.  This enlarges the class of supergroups as candidate near-horizon symmetries to others listed in Table \ref{Superconformal} of Appendix \ref{ads} that we have until now ignored. These are treated in Section \ref{new}.

\subsection{0-branes}

Similarly,  the brane-scans of Section \ref{intro} focussed  on $p\geq 1$ branes, but as discussed above $0$-branes, i.e. black holes, also have interesting $AdS_2$ near-horizon geometries also treated in Section \ref{new}.

\subsection{D-branes and AdS/CFT}

Maldacena's conjectured duality between physics in the bulk of $AdS$ and a conformal field theory on the boundary \cite{Maldacena:1997re} naturally prompts comparisons with the 
{\it membrane at the end of universe} \cite{Fifteen,BDPS,BD,BDPS2,Sutton,Duff:1999rk,Duff:1998hj,Seibergwitten,Batrachenko:2002pu}, whereby the $p$-brane occupies 
the boundary of $AdS_{p+2}$ and is described by a 
superconformal theory and to the {\it membrane/supergravity 
bootstrap}  \cite{Fifteen,BDPS,BD} which conjectured that the dynamics of the supergravity 
in the bulk of $AdS$ was dictated by the membrane on its boundary and 
vice-versa.   For example, one immediately recognizes 
that the dimensions and supersymmetries of the three conformal theories 
in Maldacena's duality are exactly the same as the boldface singleton, doubleton and tripleton 
supermultiplets of Tables \ref{scalar}, \ref{vector} and \ref{tensor}.

The main difference is that in the older work attention was focussed on {\it free} superconformal theories on the boundary as opposed to the {\it interacting} theories considered by  Maldacena.  For example, although the worldvolume 
fields of the Type IIB $3$-brane were known to be described by an 
$(n=4,d=4)$ gauge theory \cite{DLgauge}, we now know that this brane 
admits the interpretation of a Dirichlet brane \cite{Polchinski} and 
that the superposition of $N$ such branes yields a {\it non-abelian} 
$SU(N)$ gauge theory \cite{Wittenbound}. So the whole large $N$ connection was missing. Important though this omission was, it does not impair the usefulness of the near-horizon brane-scan, were it proved to be correct. The assumption that it works for all the superconformal groups in Appendix \ref{ads} (and those of other signatures too), predicts a wealth of yet more holographic duals to which we now turn. 

\section{New developments}
\la{new}

\subsection{$ \alpha'$ corrections}
\la{alpha'}

The new developments we have in mind begin with observation that higher order corrections can stretch the horizon of extremal small black holes leading to the non-singular 
$AdS_2 \times S^{D-2}$ near-horizon geometry \cite{Sen:2005kj}.

In particular for $D=5$ supergravity, coming from strings on $T^5$ or M on Calabi-Yau, the exact supersymmetric completion of the $R^2$ terms has been worked out \cite{Hanaki:2006pj} and the non-singular $AdS^2 \times S^3 \times T^5$ near-horizon geometry has been established 
\cite{Castro:2007sd,Castro:2007hc,Alishahiha:2007nn,Castro:2007ci,Cvitan:2007en,%
 Johnson,Alishahiha:2007ap}. Sure enough, the supergroups $OSp(4^*|4) \times SU(2)$  for $N=4$ and $OSp(4^*|2) \times SU(2)$ for $N=2$ make their appearance, to be compared with $SU(2|1,1) \times SU(2)$ for the large black hole \cite{Gauntlett:1998fz}.

Similarly for black strings compactified to five dimensions, one finds that higher-order corrections lead to a near-horizon geometry $OSp(4^*|4)$ and it is conjectured that similar symmetry enhancement occurs for other compactifications \cite{Dabholkar:2007gp,Lapan:2007jx,Johnson,Kraus:2007vu,Alishahiha:2008kc}. In particular,   the uncompactified $D=10$ heterotic string \cite{Dabholkar:1990yf} is expected to display a near-horizon $OSp(8|2)$  symmetry, as predicted in 1987 by the near-horizon brane-scan Table \ref{scalar}. Candidates for near-horizon supergroups of Type II strings can also be obtained by taking left and right copies of the above, again as predicted.

\subsection{The heterotic near-horizon brane-scan}
\la{16}

We now explore which near-horizon geometries are in principle allowed by the superconformal groups admitting 16 supercharges listed in Appendix \ref{16s}, where we have focussed on those admitting  bosonic symmetries $SO(p+2,2) x SO(9-k-p) x SO(k) $ listed in Table \ref{hetsubgroups} corresponding to the $AdS_{p+2} \times S^{8-k- p} \times T^k$ compactifications of Table \ref{hetgeometries}.  Following \cite{Dabholkar:2007gp,Lapan:2007jx}, there is a geometrical  $ SO(p+2,2) x SO(9-k-p)$ coming from the $AdS_{p+2} \times S^{8-k- p}$ factor and an $SO(k)$ R-symmetry coming from the $T^k$ factor.

The resulting superconformal groups are listed in the heterotic near-horizon brane -scan of Table \ref{het}.  Some caveats are in order:

1) This scan is for single fundamental branes compactified on a torus. Stacked, intersecting and/or dyonic branes with different compactifications might lead to more possibilities from Table \ref{Superconformal}.
     
2) This is just a list of {\it possibilities}; we have no idea whether they correspond to actual solutions to the $\alpha'$-corrected field equations.

3) In particular, we are not sure that one should include $AdS \times$ (flat space) cases  since these may not be  good solutions even with higher derivatives. These can easily be removed, if necessary.

4) Since we are interested here in heterotic compactifications, we have omitted those superconformal groups in \ref{16s} with right-moving supersymmetries. In particular we have no entry for strings on $AdS_3 \times S^3 \times T^4$ where one finds  $D(2,1:\alpha) \times  D(2,1:\alpha)$ in \cite{Lapan:2007jx}. The other string entries coincide with those in \cite{Lapan:2007jx}.

 5) Note that in Table \ref{hetsubgroups} there is a nice match with the extra $SU(2)$ 
 for ($N=2;  p=3,4,5$)  singletons discussed  in \cite{NST} but  $p=2$ is strange from this perspective.
 
 6) The question marks in Table \ref{het}  indicate more bosonic symmetry than  expected from Table \ref{hetsubgroups}. The meaning of this is unclear.

As a consistency check, we reproduce the $D=5$ results for both strings and black holes discussed in the previous subsection.

As another consistency check, the  heterotic superstrings at the end of $AdS_3 \times S^7$ would have to be given by the $(8_c,0)$ superconformallly invariant two-dimensional singleton field theories of the three-dimensional AdS supergroups $OSp(8|2)_c \otimes SO(2,1)$. Fortunately, this is indeed the case, as was shown by Gunaydin, Nilsson, Sierra and Townsend  \cite{Gunaydin:1986cs} in 1986.

\begin{sidewaystable}
\begin{center}
\begin{tabular}{lllllllll}
D$\uparrow$&&&&&&\\
11&.&&&&&&\\
10&.&&$AdS_3 \times S^7 $&&&&$AdS_7 \times S^3$\\
~9&.&$AdS_2 \times S^7 \times T^1$&$AdS_3 \times S^6 \times T^1$&&&$AdS_6 \times S^3 \times T^1$&$AdS_7 \times S^2 \times T^1$\\
~8&.&$AdS_2 \times S^6 \times T^2$&$AdS_3 \times S^5 \times T^2$&&$AdS_5 \times S^3 \times T^2$&$AdS_6 \times S^2 \times T^2$&$AdS_7 \times S^1 \times T^2$\\
~7&.&$AdS_2 \times S^5 \times T^3$&$AdS_3 \times S^4 \times T^3$&$AdS_4 \times S^3 \times T^3$&$AdS_5 \times S^2 \times T^3$&$AdS_6 \times S^1 \times T^3$&$AdS_7 \times T^3$\\
~6&.&$AdS_2 \times S^4 \times T^4$&$$&$AdS_4 \times S^2 \times T^4$&$AdS_5 \times S^1 \times T^4$&$AdS_6  \times T^4$&\\
~5&.&$AdS_2 \times S^3 \times T^5$&$AdS_3 \times S^2 \times T^5$&$AdS_4 \times S^1 \times T^5$&$AdS_5 \times T^5$&&\\
~4&.&$AdS_2 \times S^2 \times T^6$&$AdS_3 \times S^1 \times T^6$&$AdS_4 \times T^6$&&&\\
~3&.&$AdS_2 \times S^1 \times T^7$&$AdS_3 \times T^7$&&&&\\
~2&.&$AdS_2 \times T^8$&&&&&&\\
~1&.&&~&~&~&~&\\
~0&.&.&.&.&.&.&.\\
~~&0&1&2&3&4&5&6&d$\rightarrow$
\end{tabular}
\end{center}
\caption{The $AdS_{p+2} \times S^{8-k-p} \times T^k$ geometries for 16 supercharges}
\la{hetgeometries}
\end{sidewaystable}

\begin{sidewaystable}\centering
\begin{flushleft}
\tiny{
\begin{tabular}{lllllllll}
D$\uparrow$&&&&&&\\
11&.&&&&&&\\
10&.&&$SO(2,2)  \times SO(8) $&&&&$SO(6,2) \times SO(4)$\\
~9&.&$SO(1,2) \times SO(8)$&$SO(2,2)  \times SO(7) $&&&$SO(5,2) \times SO(4)$&$SO(6,2) \times SO(3)$\\
~8&.&$SO(1,2) \times SO(7) \times SO(2)$&$SO(2,2)  \times SO(6) \times SO(2)$&&$SO(4,2) \times SO(4) \times SO(2)$&$SO(5,2) \times SO(3) \times SO(2)$&$SO(6,2) \times SO(2) \times SO(2)$\\
~7&.&$SO(1,2) \times SO(6) \times SO(3)$&$SO(2,2)  \times SO(5) \times SO(3)$&$SO(3,2) \times SO(4) \times SO(3)$&$SO(4,2) \times SO(3) \times SO(3)$&$SO(5,2) \times SO(2) \times SO(3)$&$SO(6,2) \times SO(3)$\\
~6&.&$SO(1,2) \times SO(5) \times SO(4)$&&$SO(3,2) \times SO(3) \times SO(4)$&$SO(4,2) \times SO(2) \times SO(4) $&$SO(5,2) \times SO(4)$&\\
~5&.&$SO(1,2) \times SO(4) \times SO(5)$&$SO(2,2)  \times SO(3) \times SO(5) $&$SO(3,2) \times SO(2) \times SO(5)$&$SO(4,2) \times SO(5) $&&\\
~4&.&$SO(1,2) \times SO(3) \times SO(6)$&$SO(2,2)  \times SO(2) \times SO(6)$&$SO(3,2) \times SO(6)$&&&\\
~3&.&$SO(1,2) \times SO(2) \times SO(7)$&$SO(2,2)  \times SO(7)$&&&&\\
~2&.&$SO(1,2) \times SO(8)$&&&&&&\\
~1&.&&~&~&~&~&\\
~0&.&.&.&.&.&.&.\\
~~&0&1&2&3&4&5&6&d$\rightarrow$
\end{tabular}}
\end{flushleft}
\caption{The bosonic subgroups for 16 supercharges}
\la{hetsubgroups}
\end{sidewaystable}

\begin{sidewaystable}
\begin{center}
\begin{tabular}{lllllllll}
D$\uparrow$&&&&&&\\
11&.&~~~~~~~~~~~~~~~~~~&&&&&\\
10&.&~~~~~~~~&$OSp(8|2) \times SO(2,1)$&&&&$OSp(8^*|2) \times SO(3)$\\
~9&.&$OSp(8|2)$&$F^0(4)\times SO(2,1)$&&&$F^2(4) \times SO(3)$&$OSp(8^*|2)$\\
~8&.&$F^0(4)\times SO(2)$&$SU(4|1,1) \times SO(2,1)$&&$SU(2,2|2) \times SO(3)$&$F^2(4) \times SO(2)$&$OSp(8^*|2)  \times SO(2)?$\\
~7&.&$SU(4|1,1) \times SO(3)?$&$OSp(4^*|4) \times SO(2,1)$&$OSp(4|4)\times SO(3)$&$SU(2,2|2) \times SO(3)?$&$F^2(4) \times SO(2)$&$OSp(8^*|2)$\\
~6&.&$OSp(4^*|4) \times SO(3)$&$$&$OSp(4|4)\times SO(3)$&$SU(2,2|2) \times SO(3)$&$F^2(4) \times SO(3)$&\\
~5&.&$OSp(4^*|4) \times SO(3)$&$OSp(4^*|4) \times SO(2,1)$&$OSp(4|4)\times SO(5)?$&$SU(2,2|2) \times SO(5)?$&&\\
~4&.&$SU(4|1,1) \times SO(3)?$&$SU(4|1,1) \times SO(2,1)$&$OSp(4|4)\times SO(6)?$&&&\\
~3&.&$F^0(4)\times SO(2)$&$F^0(4) \times SO(2,1)$&&&&\\
~2&.&$OSp(8|2)$&&&&&&\\
~1&.&~~~~~~~~&~&~&~&~&\\
~0&.&.&.&.&.&.&.\\
~~&0&1&2&3&4&5&6&d$\rightarrow$
\end{tabular}
\end{center}
\caption{16 supercharges: The heterotic near-horizon brane-scan }
\la{het}
\end{sidewaystable}

 \subsection{The M/Type II near-horizon brane-scan}
\la{32}

We now explore which near-horizon geometries are in principle allowed by the superconformal groups admitting 32 supercharges listed in Appendix \ref{32s}, where we have again focussed on those admitting  bosonic symmetries $SO(p+2,2) x SO(D-1-k-p) x SO(k)$ listed in Table \ref{hetsubgroups} corresponding to the $AdS_{p+2} \times S^{D-2-k- p} \times T^k$ compactifications of Table \ref{Mgeometries}. 

The resulting superconformal groups are listed in the M/Type II near-horizon brane -scan of Table \ref{M}.  Similar caveats apply. 

Table \ref{M} has some interesting gaps, but maybe that is just  the way it is.

As a consistency check, the Type IIA and Type IIB superstrings at the end of $AdS_3 \times S^7$ would have to be given by the $(8_c,8_s)$, $(8_c,8_s)$  superconformallly invariant two-dimensional singleton field theories of the three-dimensional AdS supergroups $OSp(8|2)_c \otimes OSp(8|2)_s$ and $OSp(8|2)_c \otimes OSp(8|2)_c$, respectively. Once again, this is indeed the case \cite{Gunaydin:1986cs}.

\begin{sidewaystable}\centering
\begin{flushleft}
\begin{tabular}{ccccccccc}
D$\uparrow$&&&&&&\\
11&.&&&$AdS_4 \times S^7$&&&$AdS_7 \times S^4$\\
10&.&&$AdS_3  \times S^7 $&&$AdS_5 \times S^5$&&\\
~9&.&&$AdS_3  \times S^6 \times T^1 $&&&&\\
~8&.&&$AdS_3  \times S^5 \times T^2$&&&&\\
~7&.&&$AdS_3  \times S^4 \times T^3$&& &&\\
~6&.&&&&&&\\
~5&.&&$AdS_3  \times S^2 \times T^5$&&&&\\
~4&.&&$AdS_3  \times S^1 \times T^6$&&&&\\
~3&.&&$AdS_3 \times T^7$&&&&\\
~2&.&&&&&&&\\
~1&.&&~&~&~&~&\\
~0&.&.&.&.&.&.&.\\
~~&0&1&2&3&4&5&6&d$\rightarrow$
\end{tabular}
\end{flushleft}
\caption{The $AdS_{p+2} \times S^{8-k-p} \times T^k$ geometries for 32 supercharges}
\la{Mgeometries}
\end{sidewaystable}

\begin{sidewaystable}\centering
\begin{flushleft}
\begin{tabular}{ccccccccc}
D$\uparrow$&&&&&&\\
11&.&&&$SO(3,2) \times SO(8)$&&&$SO(6,2) \times SO(5)$\\
10&.&&$SO(2,2)  \times SO(8)^2$&&$SO(4,2) \times SO(6)$&&\\
~9&.&&$SO(2,2)  \times SO(7)^2 $&&&&\\
~8&.&&$SO(2,2)  \times SO(6)^2 \times SO(2)^2$&&&&\\
~7&.&&$SO(2,2)  \times SO(5)^2 \times SO(3)^2$&& &&\\
~6&.&&&&&&\\
~5&.&&$SO(2,2)  \times SO(3)^2 \times SO(5)^2$&&&&\\
~4&.&&$SO(2,2)  \times SO(2)^2 \times SO(6)^2$&&&&\\
~3&.&&$SO(2,2)  \times SO(7)^2$&&&&\\
~2&.&&&&&&&\\
~1&.&&~&~&~&~&\\
~0&.&.&.&.&.&.&.\\
~~&0&1&2&3&4&5&6&d$\rightarrow$
\end{tabular}
\end{flushleft}
\caption{The bosonic subgroups for 32 superchrges}
\la{subgroups32}
\end{sidewaystable}

\begin{sidewaystable}
\begin{center}
\begin{tabular}{ccccccccc}
D$\uparrow$&&&&&&\\
11&.&&&$OSp(8|4)$&&&$OSp(8^*|4)$\\
10&.&&$OSp(8|2) \times OSp(8|2)$&&$SU(2,2|4)$&&\\
~9&.&&$F^0(4)\times F^0(4)$&&&&\\
~8&.&&$SU(4|1,1) \times SU(4|1,1)$&&&&\\
~7&.&&$OSp(4^*|4) \times OSp(4^*|4)$&&&\\
~6&.&&&&&&\\
~5&.&&$OSp(4^*|4) \times OSp(4^*|4)$&&&&\\
~4&.&&$SU(4|1,1) \times SU(4|1,1) $&&&&\\
~3&.&&$F^0(4) \times F^0(4))$&&&&\\
~2&.&&&&&&&\\
~1&.&&&&&&\\
~0&.&.&.&.&.&.&.\\
~~&0&1&2&3&4&5&6&d$\rightarrow$
\end{tabular}
\end{center}
\caption{32 supercharges: The M/Type II near-horizon brane-scan }
\la{M}
\end{sidewaystable}

\section{Conclusions}

Dilatonic black holes and strings in $D=5$ display a near-horizon symmetry enhancement when $\alpha'$ corrections are taken into account. If this logic could be extended to other branes in other dimensions, it would revive the near horizon brane-scan, resolve a 21-year-old paradox and provide a wealth of new AdS/CFT dualities listed in Tables \ref{hetgeometries} to \ref{M}.

\section{Acknowledgements}

I am grateful to Mohsen Alishahiha, Hajar Ebrahim, Murat Gunaydin, Finn Larsen, Ergin Sezgin and Andy Strominger for useful conversations and correspondence.

\newpage

\appendix
\section{Superconformal groups}
\la{ads}

\begin{table}
\begin{center}
\begin{tabular}{cccc}
$d$&{\bf Supergroup}&{\bf Bosonic subgroup}&{\bf Susy}\\
\hline
6&$OSp(8^*|N)$&$SO^*(8) \times USp(N), N$ even&$8N$\\
5&$F^2(4)$&$SO(5,2) \times SU(2)$&16\\
4&$SU(2,2|N)$&$SU(2,2) \times U(N), N\neq4$&$8N$\\
~&$SU(2,2|4)$&$SU(2,2) \times SU(4)$&32\\
3&$OSp(N|4)$&$SO(N) \times Sp(4,R)$&$4N$\\
2 or 1&$OSp(N|2)$&$O(N) \times Sp(2,R)$&$2N$\\
~&$SU(N|1,1)$&$U(N) \times SU(1,1), N\neq2$&$4N$\\
~&$SU(2|1,1)$&$SU(2) \times SU(1,1)$&8\\
~&$OSp(4^*|2N)$&$SU(2) \times USp(2N) \times SU(1,1)$&$8N$\\
~&$G(3)$&$G_{2} \times SU(1,1)$&14\\
~&$F^0(4)$&$Spin(7) \times SU(1,1)$&16\\
~&$D^{}(2,1;\alpha)$&$SU(2) \times SU(2) \times SU(1,1)$&8\\
\end{tabular}
\end{center}
\caption{Superconformal groups}
\label{Superconformal}
\end{table}

Following \cite{Nahm,Gunaydin:1998km,VanProeyen:1999ni} we list the conformal supergroups and their bosonic subgroups in Table \ref{Superconformal}. 
Making use of the following isomorphisms, 
\[
SU(2) \sim SO(3) \sim USp(2)
\]
\[
SU(1,1) \sim SO(1,2) \sim Sp(2)
\]
\[
SO(4) \sim SO(3) \times SO(3)
\]
\[
SO(2,2) \sim SO(1,2) \times SO(2,1)
\]
\[
SO^*(4) \sim SO(3) \times SO(1,2)
\]
\[
USp(4) \sim SO(5)
\]
\[
Sp(4) \sim SO(3,2)
\]
\[
SU(2,2) \sim SO(4,2)
\]
we focus on those compatible with an $AdS_{p+2} \times S^{D-k- p-2} \times T^k$ geometry with either 32 or 16 supercharges. Note that the $d=2$ cases then require either a doubling or a product with a bosonic $SO(2,1)$ in order to contain the $SO(2,2)$ of $AdS_3$.  

\subsection{32 supercharges}
\la{32s}
\indent

$d=6 \qquad OSp(8^*|4) \supset SO(6,2) \times SO(5)$\\

$d=4 \qquad  SU(2,2|4) \supset SO(4,2) \times SO(6)$\\

$d=3 \qquad OSp(8|4) \supset SO(3,2) \times SO(8)$\\

$d=2 \qquad OSp(8|2) \times OSp(8|2) \supset SO(2,2) \times SO(8) \times SO(8)$\\

\subsection{16 supercharges}
\la{16s}
\indent

$d=6 \qquad OSp(8^*|2) \supset SO(6,2) \times SO(3)$\\

$d=5 \qquad F^2(4) \supset SO(5,2) \times SO(3)$\\

$d=4 \qquad SU(2,2|2) \supset SO(4,2) \times SO(3) \times U(1)$\\

$d=3 \qquad OSp(4|4) \supset SO(3,2) \times SO(3) \times SO(3)$\\

$d=2 \qquad OSp(8|2) \times SO(1,2) \supset SO(2,2) \times SO(8)$\\

$d=2 \qquad F^0(4) \times SO(1,2) \supset SO(2,2) \times SO(7)$\\

$d=2 \qquad SU(4|1,1) \times SO(1,2) \supset SO(2,2) \times SO(6) \times U(1)$\\

$d=2 \qquad OSp(4^*|4)  \times SO(1,2) \supset SO(2,2) \times SO(5) \times SO(3)$\\

$d=2 \qquad OSp(4|2) \times OSp(4|2) \supset SO(2,2) \times SO(3) \times SO(3) \times SO(3) \times SO(3)$\\ 

$d=2 \qquad OSp(5|2) \times OSp(3|2) \supset SO(2,2) \times SO(5) \times SO(3)$\\ 

$d=2 \qquad OSp(6|2) \times OSp(2|2) \supset SO(2,2) \times SO(6) \times U(1)$\\ 

$d=2 \qquad OSp(7|2) \times OSp(1|2) \supset SO(2,2) \times SO(7)$\\ 

$d=2 \qquad D(2,1:\alpha) \times  D(2,1:\alpha) \supset SO(2,2) \times SO(3) \times SO(3)\times SO(3) \times SO(3)$\\
 
$d=2 \qquad SU(2|1,1) \times SU(2|1,1) \supset SO(2,2) \times SO(3) \times SO(3)$\\

$d=2 \qquad OSp(4^*|2) \times OSp(4^*|2) \supset SO(2,2)  \times SO(3) \times SO(3)\times SO(3) \times SO(3)$\\

$d=1\qquad OSp(8|2) \supset SO(1,2) \times SO(8)$\\

$d=1 \qquad F^0(4) \supset SO(1,2) \times SO(7)$\\

$d=1 \qquad SU(4|1,1) \supset SO(1,2) \times SO(6) \times U(1)$\\

$d=1 \qquad OSp(4^*|4) \supset SO(1,2) \times SO(5) \times SO(3)$\\


\newpage
\begin{thebibliography}{99}

\bibitem{AETW}
A.~Achucarro, J.~Evans, P.~Townsend and D.~Wiltshire,
\newblock {\sl Super p-branes},
\newblock Phys. Lett. {\bf B198} (1987) 441.

\bibitem{Green:1983wt}
  M.~B.~Green and J.~H.~Schwarz,
  ``Covariant Description Of Superstrings,''
  Phys.\ Lett.\  B {\bf 136} (1984) 367.

\bibitem{Duff:1987bx}
  M.~J.~Duff, P.~S.~Howe, T.~Inami and K.~S.~Stelle,
  ``Superstrings in D = 10 from supermembranes in D = 11,''
  Phys.\ Lett.\  B {\bf 191}, 70 (1987).


\bibitem{BD}
M. P. Blencowe and M. J. Duff,
{\sl Supersingletons},
Phys. Lett. {\bf B203}, 229 (1988).

\bibitem{Fifteen}
M. J. Duff,
{\sl Supermembranes:  The first fifteen weeks},
 Class. Quant. Grav. {\bf 5}, 189 (1988).

 \bibitem{BDPS}
E. Bergshoeff, M. J. Duff, C. N. Pope and E. Sezgin,
{\sl Supersymmetric supermembrane vacua and singletons},
 Phys. Lett. {\bf B199}, 69 (1988).
 
 \bibitem{Nahm}
W. Nahm,
{\sl Supersymmetries and their representations},
Nucl. Phys. {\bf B135} (1978) 149.

  \bibitem{Gunaydin1}
M. G\"{u}naydin,
{\sl Singleton and doubleton supermultiplets of space-time supergroups 
and infinite spin superalgebras}, in Proceedings of the 1989 Trieste 
Conference ``Supermembranes and Physics in $2+1$ Dimensions'', 
eds Duff, Pope and Sezgin, World Scientific 1990.

\bibitem{Gunaydin:1986cs}
  M.~Gunaydin, B.~E.~W.~Nilsson, G.~Sierra and P.~K.~Townsend,
  ``Singletons And Superstrings,''
  Phys.\ Lett.\  B {\bf 176} (1986) 45.

\bibitem{NST}
H. Nicolai, E. Sezgin and Y. Tanii,
{\sl Conformally invariant supersymmetric field theories on $S^{p}
\times S^{1}$ and super $p$-branes},
Nucl. Phys. {\bf B305} (1988) 483.

 \bibitem{DTV}
M. J. Duff, P. K. Townsend and P. van Nieuwenhuizen,
{\sl Spontaneous compactification of supergravity on the three-sphere},
Phys. Lett. {\bf B122} (1983) 232.

\bibitem{GT}
G. W. Gibbons and P. K. Townsend,
{\sl Vacuum interpolation in supergravity via super $p$-branes},
Phys. Rev. Lett. {\bf 71} (1993) 3754.


 \bibitem{DGT}
M. J. Duff, G. W. Gibbons and P. K. Townsend,
{\sl Macroscopic superstrings as interpolating solitons},
Phys. Lett. B. {\bf 332} (1994) 321.



\bibitem{Dabholkar:2007gp}
  A.~Dabholkar and S.~Murthy,
  ``Fundamental Superstrings as Holograms,''
  JHEP {\bf 0802}, 034 (2008)
  [arXiv:0707.3818 [hep-th]].
  
\bibitem{Lapan:2007jx}
  J.~M.~Lapan, A.~Simons and A.~Strominger,
  ``Nearing the Horizon of a Heterotic String,''
  arXiv:0708.0016 [hep-th].
  
\bibitem{Dabholkar:1990yf}
  A.~Dabholkar, G.~W.~Gibbons, J.~A.~Harvey and F.~Ruiz Ruiz,
  ``SUPERSTRINGS AND SOLITONS,''
  Nucl.\ Phys.\  B {\bf 340}, 33 (1990).

  \bibitem{Johnson}
  C. V. Johnson, 
  ÒHeterotic coset models of microscopic strings and black holes,Ó 
arXiv:0707.4303[hep-th].
  
\bibitem{Kraus:2007vu}
  P.~Kraus, F.~Larsen and A.~Shah,
  ``Fundamental Strings, Holography, and Nonlinear Superconformal Algebras,''
  JHEP {\bf 0711}, 028 (2007)
  [arXiv:0708.1001 [hep-th]].

  \bibitem{String}
M.~J. Duff, R.~R. Khuri and J.~X. Lu,
{\sl String solitons},
Phys. Rep. {\bf 259} (1995) 213, hep-th/9412184


\bibitem{Duff:1990xz}
  M.~J.~Duff and K.~S.~Stelle,
  ``Multi-membrane solutions of D = 11 supergravity,''
  Phys.\ Lett.\  B {\bf 253}, 113 (1991).

\bm{lublack} 
M.J. Duff and J.X. Lu, 
{\sl Black and super $p$-branes in diverse dimensions}, 
Nucl. Phys. {\bf B416} (1994) 301. 

\bibitem{DR2}
M.~J. Duff and J.~Rahmfeld,
\newblock {\sl Massive string states as extreme black holes},
\newblock Phys. Lett. {\bf B345} (1995) 441.

\bibitem{DR}
M.J. Duff and J. Rahmfeld, 
{\sl Bound states of black holes and other $p$-branes},
Nucl. Phys. {\bf B481} (1996) 332.

\bibitem{KKLP}
N. Khviengia, Z. Khviengia, H. L\"{u} and C.N. Pope,
{\sl Interacting $M$-branes and bound states},
Phys. Lett. {\bf B388} (1996) 21. 

\bm{Rahmfeld} M. J. Duff, S. Ferrara, R. Khuri and J. Rahmfeld,
{\sl Supersymmetry and dual string solitons}, Phys. Lett. {\bf B356} 
(1995) 479.

\bibitem{T} A. A. Tseytlin, {\sl Extreme dyonic black holes in string theory}, 
Mod. Phys. Lett. {\bf A11} (1996) 689, hep-th/9601177.


\bm{CT1} M. Cvetic and A. A. Tseytlin, {\sl General class of BPS 
saturated black holes as exact superstring solutions}, Phys. Lett. 
{\bf B366} (1996) 95, hep-th/9510097.

\bm{CT2} M. Cvetic and A. A. Tseytlin, {\sl Solitonic strings and BPS 
saturated dyonic black holes}, Phys. Rev. {\bf D53} (1996) 5619, 
hep-th/9512031.
  
\bibitem{FG}
D. Z. Freedman and G. W. Gibbons,
{\sl Electrovac ground state in gauged $SU(2) \times SU(2)$ 
supergravity},
Nucl. Phys. {\bf B233} (1984) 24.

\bibitem{DLgauge}
M. J. Duff and J. X. Lu,
\newblock {\sl The self-dual Type IIB super 3-brane},
\newblock Phys. Lett. {\bf B273} (1991) 409.

\bibitem{CHS1}
C. G.~Callan, J.~A. Harvey and A.~Strominger,
\newblock {\sl World sheet approach to heterotic instantons and solitons},
\newblock Nucl. Phys. {\bf  B359} (1991) 611.

\bibitem{CHS2}
C. G.~Callan, J.~A. Harvey and A.~Strominger,
\newblock {\sl Worldbrane actions for string solitons},
\newblock Nucl. Phys. {\bf  B367} (1991) 60.


\bibitem{HS}
G.~T. Horowitz and A.~Strominger,
\newblock {\sl Black strings and p-branes},
\newblock Nucl. Phys. {\bf B360} (1991) 197.


\bibitem{Luscan}
M. J. Duff and J. X. Lu,
\newblock{\sl Type II $p$-branes: the brane-scan revisited},
\newblock Nucl. Phys. {\bf B390} (1993) 276.

\bibitem{Duff:1991sz}
  M.~J.~Duff and J.~X.~Lu,
  ``A Duality between strings and five-branes,''
  Class.\ Quant.\ Grav.\  {\bf 9}, 1 (1992).

\bibitem{super}  
M. J. Duff, 
{\sl Supermembranes}, lectures given at 
the T. A. S. I. Summer School, University of Colorado, Boulder, June 1996;
the Topical Meeting, Imperial College, London, July 1996 and the 26th
British Universities Summer School in Theoretical Elementary Particle
Physics, University of Swansea, September 1996.
Int. J. Mod. Phys. {\bf A11} (1996) 5623, hep-th/9611203. 



\bm{Gueven}
R. Gueven,
{\sl Black $p$-brane solutions of $D = 11$ supergravity theory},
Phys. Lett. {\bf B276} (1992) 49.


\bibitem{Duff:1999rk}
  M.~J.~Duff,
  ``Lectures on branes, black holes and anti-de Sitter space,''
  arXiv:hep-th/9912164.
  
  
\bibitem{Duff:1998hj}
  M.~J.~Duff,
  ``Anti-de Sitter space, branes, singletons, superconformal field theories
  and all that,''
  Int.\ J.\ Mod.\ Phys.\  A {\bf 14}, 815 (1999)
  [arXiv:hep-th/9808100].
  

\bibitem{BD2}
M. P. Blencowe and M. J. Duff,
{\sl Supermembranes and the signature of spacetime},
 Nucl. Phys. {\bf B310} 387 (1988).


\bibitem{Hull}
C. M. Hull,
{\sl Duality and the signature of spacetime},
hep-th/9807127.

\bibitem{BDPS2}
E. Bergshoeff, M. J. Duff, C. N. Pope and E. Sezgin,
{\sl Compactifications of the eleven-dimensional supermembrane},
Phys. Lett. {\bf B224}, 71 (1989).

\bibitem{Sutton}
M. J. Duff and C. Sutton,
{\sl The membrane at the end of the universe}
New. Sci. {\bf 118} (1988) 67.

\bibitem{Seibergwitten}
N. Seiberg and E. Witten,
{\sl The D1/D5 system and singular CFT},
hep-th/9903224.


\bibitem{Batrachenko:2002pu}
  A.~Batrachenko, M.~J.~Duff and J.~X.~Lu,
  ``The membrane at the end of the (de Sitter) universe,''
  Nucl.\ Phys.\  B {\bf 762}, 95 (2007)
  [arXiv:hep-th/0212186].


\bibitem{Maldacena:1997re}
  J.~M.~Maldacena,
  ``The large N limit of superconformal field theories and supergravity,''
  Adv.\ Theor.\ Math.\ Phys.\  {\bf 2}, 231 (1998)
  [Int.\ J.\ Theor.\ Phys.\  {\bf 38}, 1113 (1999)]
  [arXiv:hep-th/9711200].

\bibitem{Polchinski}
J. Polchinski,
{\sl Dirichlet branes and Ramond-Ramond charges},
Phys. Rev. Lett. {\bf 75} (1995) 4724

\bm{Wittenbound} 
E. Witten, 
{\sl Bound states of strings and $p$-branes}, 
Nucl. Phys. {\bf B460} (1996) 335, hep-th/9510135.

\bibitem{Sen:2005kj}
  A.~Sen,
  ``Stretching the horizon of a higher dimensional small black hole,''
  JHEP {\bf 0507}, 073 (2005)
  [arXiv:hep-th/0505122].
  
\bibitem{Hanaki:2006pj}
  K.~Hanaki, K.~Ohashi and Y.~Tachikawa,
  ``Supersymmetric Completion of an $R^2$ Term in Five-Dimensional Supergravity,''
  Prog.\ Theor.\ Phys.\  {\bf 117}, 533 (2007)
  [arXiv:hep-th/0611329].
  
\bibitem{Castro:2007sd}
  A.~Castro, J.~L.~Davis, P.~Kraus and F.~Larsen,
  ``5D attractors with higher derivatives,''
  JHEP {\bf 0704}, 091 (2007)
  [arXiv:hep-th/0702072].
  
\bibitem{Castro:2007hc}
  A.~Castro, J.~L.~Davis, P.~Kraus and F.~Larsen,
  ``5D Black Holes and Strings with Higher Derivatives,''
  JHEP {\bf 0706}, 007 (2007)
  [arXiv:hep-th/0703087].
 
\bibitem{Alishahiha:2007nn}
  M.~Alishahiha,
  ``On R**2 corrections for 5D black holes,''
  JHEP {\bf 0708}, 094 (2007)
  [arXiv:hep-th/0703099].


\bibitem{Castro:2007ci}
  A.~Castro, J.~L.~Davis, P.~Kraus and F.~Larsen,
  ``Precision entropy of spinning black holes,''
  JHEP {\bf 0709}, 003 (2007)
  [arXiv:0705.1847 [hep-th]].
  

  
\bibitem{Cvitan:2007en}
  M.~Cvitan, P.~D.~Prester, S.~Pallua and I.~Smolic,
  ``Extremal black holes in D=5: SUSY vs. Gauss-Bonnet corrections,''
  JHEP {\bf 0711}, 043 (2007)
  [arXiv:0706.1167 [hep-th]].
  
\bibitem{Alishahiha:2007ap}
  M.~Alishahiha, F.~Ardalan, H.~Ebrahim and S.~Mukhopadhyay,
  ``On 5D Small Black Holes,''
  JHEP {\bf 0803}, 074 (2008)
  [arXiv:0712.4070 [hep-th]].
  
\bibitem{Gauntlett:1998fz}
  J.~P.~Gauntlett, R.~C.~Myers and P.~K.~Townsend,
 ``Black holes of D = 5 supergravity,''
  Class.\ Quant.\ Grav.\  {\bf 16}, 1 (1999)
  [arXiv:hep-th/9810204].
  
\bibitem{Alishahiha:2008kc}
  M.~Alishahiha and S.~Mukhopadhyay,
  ``On Six Dimensional Fundamental Superstrings as Holograms,''
  arXiv:0803.0685 [hep-th].
  
\bibitem{Gunaydin:1998km}
  M.~Gunaydin and D.~Minic,
  ``Singletons, doubletons and M-theory,''
  Nucl.\ Phys.\  B {\bf 523}, 145 (1998)
  [arXiv:hep-th/9802047].
  
\bibitem{VanProeyen:1999ni}
  A.~Van Proeyen,
  ``Tools for supersymmetry,''
  arXiv:hep-th/9910030.
  
  
\end{thebibliography}
\end{document}